\documentclass{article} % For LaTeX2e
\usepackage{iclr2027_conference,times}

\usepackage{amsmath,amsfonts,bm}

\def\eqref#1{equation~\ref{#1}}
\def\1{\bm{1}}

\DeclareMathAlphabet{\mathsfit}{\encodingdefault}{\sfdefault}{m}{sl}
\SetMathAlphabet{\mathsfit}{bold}{\encodingdefault}{\sfdefault}{bx}{n}

\usepackage{amsmath,amssymb}
\usepackage{booktabs}
\usepackage{graphicx}
\usepackage{array}
\usepackage{multirow}
\usepackage{tabularx}
\usepackage{enumitem}
\usepackage{microtype}
\usepackage{xcolor}
\usepackage{hyperref}
\hypersetup{
    hidelinks
}
\usepackage{adjustbox}

\usepackage{url}
\usepackage{wrapfig}
\newlength{\sideTableWidth}
\newcommand{\compacttable}{%
  \scriptsize
  \setlength{\tabcolsep}{3.2pt}%
  \renewcommand{\arraystretch}{0.92}%
}

\makeatother
\title{LIFT: A Lifecycle-aware Interaction \\ Factorization Transformer for \\ Unified Retrieval and Ranking}

\author{Keji Miao \\
Beijing University of Post and Telecommunications\\
Beijing, China \\
\texttt{corgi\_m@bupt.edu.cn}
\And
Enhao Cheng \thanks{Keji Miao and Enhao Cheng contributed equally.} \\
Southeast University\\
Nanjing, China \\
\texttt{enhaocheng@seu.edu.cn} \\
\And
Yan Li \\
Xiaohongshu\\
Beijing, China
\And
Qun Li \\
Independent\\
Beijing, China
\And
Qi Zhang \\
Beijing University of Post and Telecommunications\\
Beijing, China
\And
Jie Yuan\thanks{Corresponding Author.} , Xiaoyong Li \\
Beijing University of Post and Telecommunications\\
Beijing, China \\
\texttt{\{yuanjie, lixiaoyong\}@bupt.edu.cn}
} 

\iclrfinalcopy % Uncomment for camera-ready version, but NOT for submission.
\begin{document}

\maketitle
\fancyhead{}
\lhead{\textit{Preprint}}
\rhead{}
\begin{abstract}
Cascaded recommender systems use the same user history for retrieval and
ranking, while the two stages have access to different information at different points of an interaction. We propose the Lifecycle-aware
Interaction Factorization Transformer (LIFT), which decomposes each
interaction into ordered Request, Item, Context, and Action states and
models them as a causal sequence. Retrieval reads the Request state, while
ranking reads the Context state, allowing both tasks to share history
modeling while preserving stage-specific information. LIFT instantiates
this representation with Role-Conditioned Attention and a lightweight
Pre-LN Bias. On ML-20M and Taobao, LIFT achieves the highest Joint Score
among the evaluated joint models, improving over the strongest baselines
by 4.9\% and 3.6\%, respectively. Loss-weight sweeps show favorable
retrieval--ranking trade-offs, while ablations and scaling analyses further examine lifecycle sequence
construction, model components, and capacity settings.
\end{abstract}
\section{Introduction}

Modern recommender systems commonly use a retrieval--ranking cascade \citep{qin2022rankflow,liu2022neural}.
Retrieval selects candidates from a large item collection \citep{gao2020deep,zhu2018learning}, while ranking
predicts user feedback \citep{zhou2019deep,lian2018xdeepfm} using the retrieved candidates and additional
contextual features. Both stages depend on the same user history, creating substantial
opportunities for shared historical modeling \citep{gao2020deep}. Recent unified architectures reuse history representations across tasks
and recommendation stages through shared modeling, joint training, and
cross-stage caching
\citep{deng2025onerec,dai2026onepiece,li2026unipinrec}, yet retrieval and
ranking operate with different information available within an interaction.

Sharing history computation alone does not resolve the representational
difference between retrieval and ranking. Retrieval makes a decision from
the user history and the current request, whereas ranking operates after a
candidate item has been introduced and can additionally use candidate-side
context. Sequential models, however, commonly represent each interaction as an item- or event-level state \citep{hidasi2015session,kang2018self,zhai2024actions}. This leaves
the information-availability boundaries within an interaction implicit.
A unified retrieval--ranking model therefore benefits from representing
the intermediate states through which an interaction evolves.

\begin{wrapfigure}{r}{0.48\linewidth}
    \vspace{-1\baselineskip}
    \centering
    \includegraphics[width=\linewidth]{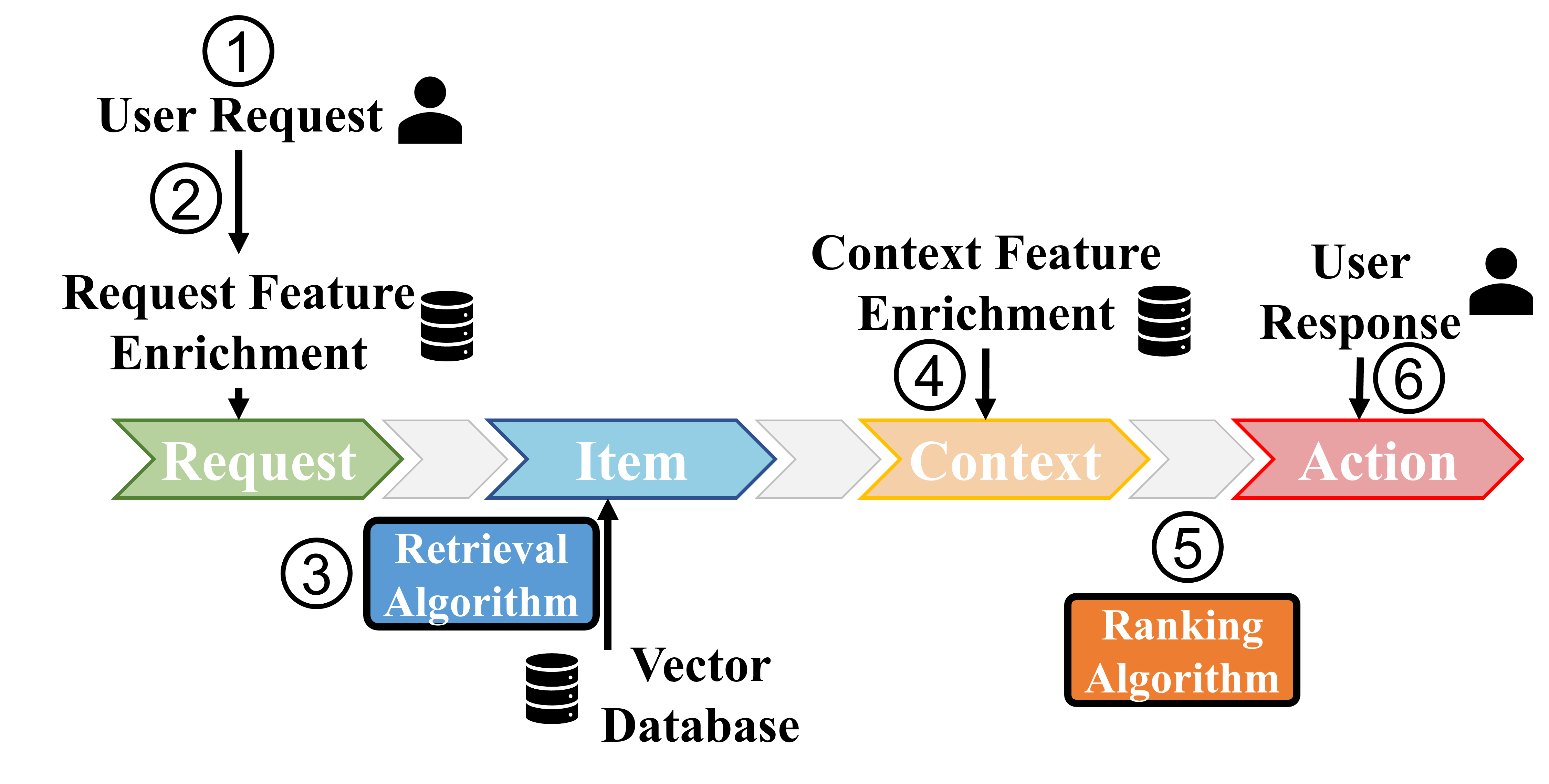}
    \vspace{-2\baselineskip}
    \caption{Lifecycle of a recommendation interaction.}
    \vspace{-1\baselineskip}
    \label{fig:lifecycle}
\end{wrapfigure}

Figure~\ref{fig:lifecycle} illustrates the interaction lifecycle considered in this work. A Request is linked to an Item through retrieval; contextual information is then incorporated, and ranking predicts the resulting user Action. The arrows indicate the logical order in which information becomes available during an interaction, while the four roles together form a single interaction record.

Following this progression, we propose the Lifecycle-aware Interaction
Factorization Transformer (LIFT). LIFT factorizes each interaction into
four ordered states---Request, Item, Context, and Action---according to
when information becomes available, and serializes them into a causal
sequence. Retrieval reads the Request state, while ranking reads the
Context state after the candidate and its contextual features have become
available.

The serialized lifecycle sequence also supports shared historical
computation across recommendation stages. During online serving, the
historical sequence is prefilled once and its cached states are reused by
retrieval and ranking, which read the Request and Context states,
respectively.

We instantiate this representation with Role-Conditioned Attention, which
shares Query and Key projections while conditioning Value and Output
transformations on lifecycle roles. A lightweight Pre-LN Bias further
conditions the feed-forward path on the current role. Retrieval and ranking
retain their task-specific candidate processing, supervision, and prediction
heads while sharing the causal history encoder.

We compare LIFT with single-task models, multi-task models, and unified sequence models on ML-20M and Taobao. LIFT achieves the highest Joint Score among the evaluated joint models on both datasets. On ML-20M, it obtains a score of 0.2724, compared with 0.2597 for the strongest baseline, HSTU, a relative improvement of 4.9\%. On Taobao, it improves over the strongest baseline by 3.6\%. Loss-weight sweeps show a more favorable retrieval--ranking quality trade-off for LIFT over the evaluated range. Ablation studies further examine lifecycle sequence construction, Role-Conditioned Attention, and Pre-LN Bias.

Our contributions are summarized as follows:
\begin{itemize}
    \item We propose lifecycle-aware interaction factorization, representing each interaction as ordered Request, Item, Context, and Action states and assigning stage-aligned readouts to retrieval and ranking.
    \item We instantiate lifecycle-aware interaction factorization with a causal history encoder that uses Role-Conditioned Attention and a lightweight additive Pre-LN Bias, while preserving the task-specific interfaces of retrieval and ranking.
    \item We conduct controlled experiments on ML-20M and Taobao, evaluating joint recommendation quality, retrieval--ranking trade-offs, and the effects of lifecycle design choices and model components.
\end{itemize}

\section{Related Work}
\subsection{Sequential Recommendation Modeling}
Sequential recommendation organizes user behavior as a temporally ordered
interaction sequence to model behavioral order, evolving interests, and
long-range dependencies \citep{quadrana2018sequence, pan2026survey,zhou2019deep}. GRU4Rec encodes behavior sequences with recurrent
neural networks, while SASRec introduces self-attention for sequential
recommendation \citep{hidasi2015session,kang2018self}. Subsequent work
extends Transformer-based sequence modeling through bidirectional
representations, temporal intervals, and self-supervised objectives
\citep{sun2019bert4rec,li2020time,zhou2020s3}. HSTU further
formulates recommendation as sequence transduction for large-scale,
non-stationary behavior streams \citep{zhai2024actions}. LIFT represents the request as a distinct token, enabling the model to make explicit use of request-side information that is available at retrieval time.
\subsection{Unified Modeling of Retrieval and Ranking}
Prior work has explored different levels of integration between retrieval
and ranking \citep{qin2022rankflow}. CoRR jointly improves a retriever and a ranker through adaptive
negative sampling and knowledge distillation
\citep{huang2023cooperative}. More recent work moves toward unified model
architectures. UniGRF reformulates retrieval and ranking as generative
objectives within a common framework \citep{zhang2025killing}. OneRec
replaces the cascaded pipeline with a generative recommender and iterative
preference alignment \citep{deng2025onerec}. OnePiece uses structured
context for both retrieval and ranking, together with block-wise latent
reasoning and progressive multi-task training \citep{dai2026onepiece}.
UniPinRec unifies the input format, model architecture, and training
pipeline, and reuses user-history computation across stages through
cross-stage KV caching \citep{li2026unipinrec}.
LIFT explores a different representation granularity by factorizing each
interaction according to information availability. Existing sequential
formulations typically associate an interaction or action event with a
single sequence state, whereas LIFT exposes multiple causal states within
the interaction itself. Their order follows when Request, Item, Context,
and Action information becomes available, allowing retrieval and ranking
to read distinct lifecycle states within the same causal history model.
\subsection{Multi-Task Learning and Structured Parameter Sharing}
Multi-task learning coordinates parameter use across multiple prediction
tasks through shared representations, gated experts, or a combination of
shared and task-specific experts \citep{ruder2017overview}. Shared-bottom parameter sharing follows
the classical multi-task learning paradigm \citep{caruana1997multitask}.
MMoE uses task-specific gating over shared experts
\citep{ma2018modeling}, while PLE further organizes shared and
task-specific experts \citep{tang2020progressive}. LIFT combines shared history encoding with stage-aligned state readouts: retrieval and ranking reuse the same causal interaction sequence while reading different states suited to their respective stages.

\section{Method}
\subsection{Problem Setup and System Overview}
\label{sec:overview}
In LIFT, the retrieval task generates candidate items based on the user history and the current request, while the ranking task predicts user feedback for candidate items. During training, both tasks share the same lifecycle sequence and historical encoding results, while differing in state readout positions, candidate processing, negative sampling, and supervision. During online serving, the ranking stage further introduces candidate-related features after the retrieval candidates have been determined. LIFT unifies the two tasks within a common lifecycle-based sequence modeling framework.

\begin{figure}[h]
    \vspace{-1\baselineskip}
    \centering
    \includegraphics[width=0.98\linewidth]{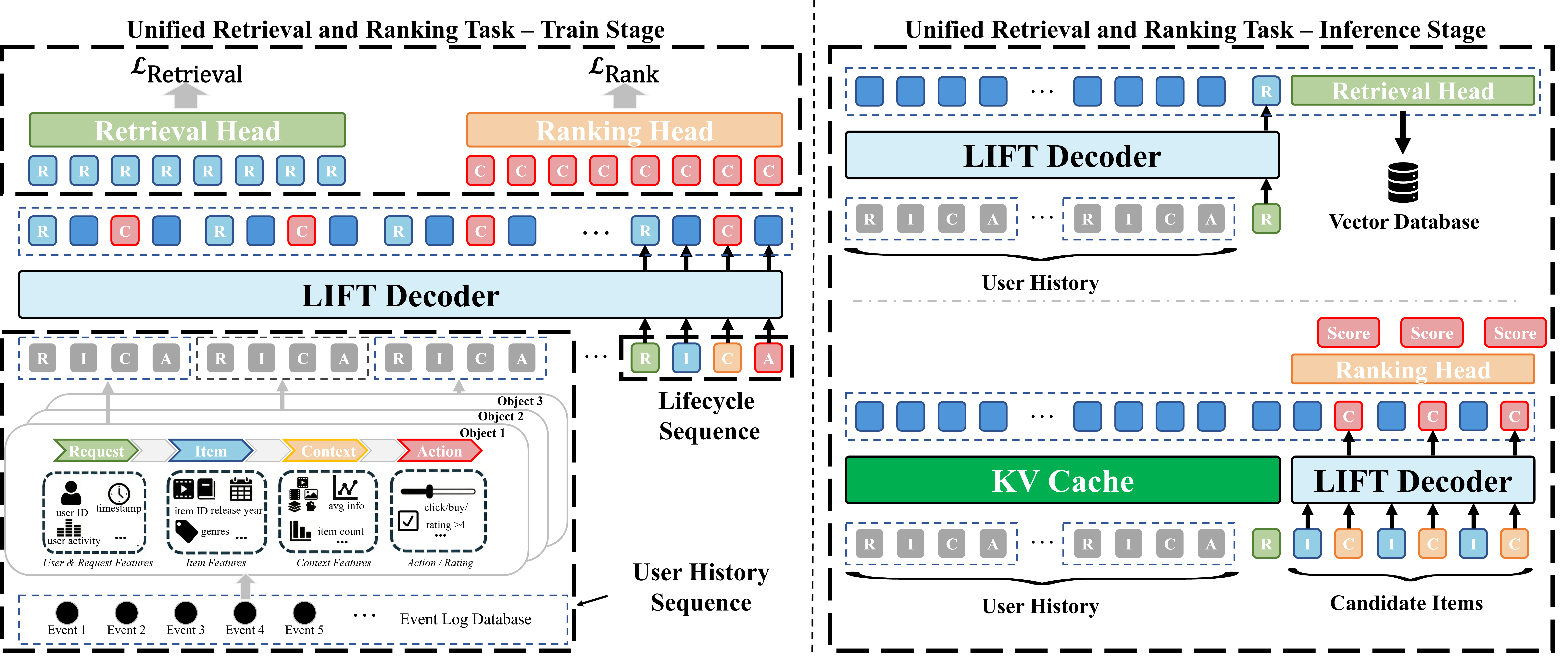}
    \vspace{-1\baselineskip}
    \caption{Overview of LIFT's unified training and online serving pipelines.}
    \vspace{-1\baselineskip}
    \label{fig:overview}
\end{figure}

Figure~\ref{fig:overview} summarizes the overall architecture and information flow of LIFT, including the shared lifecycle sequence and the LIFT Decoder, the Request- and Context-based readout branches, and the online retrieval--ranking cascade.

LIFT organizes user interactions according to four lifecycle roles: Request, Item, Context, and Action. For a user $u$ with an interaction sequence of length $T$, let $\Pi_{\mathrm{Req}}$ and $\Pi_{\mathrm{Ctx}}$ denote the readout operators for the Request and Context states, respectively. The unified encoding and task prediction process is given by:
\begin{equation}
\begin{aligned}
\mathbf{Z}_{u,\leq T}
&=
\left[
\mathbf{z}^{R}_{u,1},
\mathbf{z}^{I}_{u,1},
\mathbf{z}^{C}_{u,1},
\mathbf{z}^{A}_{u,1},
\ldots,
\mathbf{z}^{R}_{u,T},
\mathbf{z}^{I}_{u,T},
\mathbf{z}^{C}_{u,T},
\mathbf{z}^{A}_{u,T}
\right],
\\
\mathbf{H}_{u,\leq T}
&=
F_{\theta}
\left(
\mathbf{Z}_{u,\leq T}
\right),
\label{eq:lift_forward}
\end{aligned}
\end{equation}
\begin{equation}
\begin{aligned}
\widehat{\mathbf{y}}_{u,\leq T}^{\mathrm{ret}}
=
g_{\mathrm{ret}}
\left(
\Pi_{\mathrm{Req}}
\left(
\mathbf{H}_{u,\leq T}
\right)
\right), \quad
\widehat{\mathbf{y}}_{u,\leq T}^{\mathrm{rank}}
=
g_{\mathrm{rank}}
\left(
\Pi_{\mathrm{Ctx}}
\left(
\mathbf{H}_{u,\leq T}
\right)
\right).
\end{aligned}
\end{equation}
Here, $\mathbf{z}^{R}$, $\mathbf{z}^{I}$, $\mathbf{z}^{C}$, and
$\mathbf{z}^{A}$ denote Request, Item, Context, and Action token representations,
respectively. The lifecycle-aware representation is described in Section~\ref{sec:lifecycle_representation}.

In Eq.~(\ref{eq:lift_forward}), $F_{\theta}$ denotes the shared causal
Transformer Decoder \citep{vaswani2017attention}, and $\mathbf{H}_{u,\leq T}$ denotes the hidden-state sequence aligned with the lifecycle tokens. The outputs $\widehat{\mathbf{y}}_{u,\leq T}^{\mathrm{ret}}$ and $\widehat{\mathbf{y}}_{u,\leq T}^{\mathrm{rank}}$ denote the vectorized candidate-level outputs of the retrieval and ranking tasks at valid temporal positions, respectively. The function $g_{\mathrm{ret}}$ maps Request states to outputs over the retrieval candidate space, while $g_{\mathrm{rank}}$ maps Context states to outputs over the ranking candidate set. The Decoder uses shared parameters to model dependencies across historical interactions while retaining role-dependent local state transformations. The detailed interaction modeling mechanism is presented in Section~\ref{sec:role_conditioned_attention}.

During training, the two tasks produce supervision through separate task heads and loss functions. The shared LIFT Decoder is jointly optimized by both tasks, while candidate construction, negative sampling, and supervision labels remain task-specific:
\begin{equation}
\begin{gathered}
\mathcal{L}_{\mathrm{ret}}
=
\ell_{\mathrm{ret}}
\left(
\widehat{\mathbf{y}}_{u,\leq T}^{\mathrm{ret}},
\mathbf{y}_{u,\leq T}^{\mathrm{ret}},
\mathcal{N}^{\mathrm{ret}}
\right),
\quad
\mathcal{L}_{\mathrm{rank}}
=
\ell_{\mathrm{rank}}
\left(
\widehat{\mathbf{y}}_{u,\leq T}^{\mathrm{rank}},
\mathbf{y}_{u,\leq T}^{\mathrm{rank}},
\mathcal{N}^{\mathrm{rank}}
\right).
\\
\mathcal{L}
=
\lambda_{\mathrm{ret}}\mathcal{L}_{\mathrm{ret}}
+
\lambda_{\mathrm{rank}}\mathcal{L}_{\mathrm{rank}},
\end{gathered}
\label{eq:task_training}
\end{equation}
Here, $\mathbf{y}_{u,\leq T}^{\mathrm{ret}}$ denotes the target-item labels, while $\mathbf{y}_{u,\leq T}^{\mathrm{rank}}$ denotes the ranking labels constructed from the corresponding Action feedback. $\mathcal{N}^{\mathrm{ret}}$ and $\mathcal{N}^{\mathrm{rank}}$ represent the negative samples or candidate-sampling information used by the two tasks, respectively. The functions $\ell_{\mathrm{ret}}$ and $\ell_{\mathrm{rank}}$ compute and aggregate the task losses over valid temporal positions according to their respective supervision and sampling strategies.

During online serving at time $t$, LIFT first performs a single shared prefill over the user's historical lifecycle sequence and caches the resulting Key--Value (KV) states as $\mathrm{KV}_{u,<t}$. The retrieval stage reads the Request state, computes scores for candidate items, and generates a candidate set. The ranking stage then constructs the corresponding Item and Context inputs from the retrieved candidates and computes their ranking scores.
\begin{equation}
\begin{aligned}
\widehat{\mathcal{C}}_{u,t}^{\mathrm{ret}}
&=
\operatorname{RetrieveByTopK}
\left(
\mathrm{KV}_{u,<t};
\widetilde{\mathbf{z}}^{R}_{u,t};
\mathcal{V}_{t}^{\mathrm{ret}}
\right),
\\
\widehat{\mathcal{S}}_{u,t}^{\mathrm{rank}}
&=
\operatorname{RankingByScore}
\left(
\mathrm{KV}_{u,<t};
\widetilde{\mathbf{z}}^{R}_{u,t};
\left\{
(\widetilde{\mathbf{z}}^{I}_{u,t,c},
\widetilde{\mathbf{z}}^{C}_{u,t,c})
\mid
c \in \widehat{\mathcal{C}}_{u,t}^{\mathrm{ret}}
\right\}
\right).
\end{aligned}
\label{eq:online_serving}
\end{equation}
Here, the tilde denotes representations constructed from the current request
or candidate-specific features, in contrast to historical lifecycle tokens. $\operatorname{RetrieveByTopK}(\cdot)$ selects the $K$ highest-scoring candidate items from the candidate item space $\mathcal{V}_{t}^{\mathrm{ret}}$ based on the retrieval scores, forming the input set $\widehat{\mathcal{C}}_{u,t}^{\mathrm{ret}}$ for the ranking stage. The ranking stage constructs the corresponding Item and Context tokens for this candidate set and computes ranking scores. Finally, $\operatorname{RankingByScore}(\cdot)$ reorders the candidates according to their scores to produce the ordered candidate-item sequence $\widehat{\mathcal{S}}_{u,t}^{\mathrm{rank}}$. This process allows the two tasks to share historical dependency modeling while preserving their respective candidate-processing and prediction interfaces.

\subsection{Lifecycle-aware Interaction Representation}
\label{sec:lifecycle_representation}
As illustrated in Figure~\ref{fig:overview}, a user interaction contains information with different semantics, including a request, an item, contextual features, and user feedback. Based on the semantics of these fields and their availability during the recommendation process, we represent the interaction of user $u$ at time $t$ as
\begin{equation}
e_{u,t}
=
\left(
x^{R}_{u,t},
x^{I}_{u,t},
x^{C}_{u,t},
x^{A}_{u,t}
\right),
\label{eq:interaction_decomposition}
\end{equation}
where the four types of fields describe the trigger condition, interaction object, candidate context, and user response, respectively. This decomposition is determined by the generation dependencies and availability of the fields, independently of how the fields are organized in a particular data table.
\begin{itemize}[leftmargin=1.5em,itemsep=0.15em,topsep=0.2em]
\item $x^{R}_{u,t}$ denotes the Request information, namely the request-side inputs available before the candidate items are determined. It includes the user identifier, current query or request features, session information, temporal features, and device- or traffic-related context, which together describe the conditions that trigger the interaction \citep{wang2018billion,zhu2018learning,gao2020deep}.
\item $x^{I}_{u,t}$ denotes the Item information, namely the relatively stable attributes of the interacted item, including its identifier, category, brand, and content features. In large-scale retrieval systems, item-side representations are commonly
precomputed or organized into retrieval structures for efficient candidate
generation \citep{wang2018billion,gao2020deep}.
\item $x^{C}_{u,t}$ denotes the Context information, namely the candidate-related features formed after the item is introduced into the current request and interaction scenario. Examples include time-windowed item statistics, user--item interaction statistics, candidate sources, and scene features determined before scoring \citep{cheng2016wide,zhou2019deep,lian2018xdeepfm,song2019autoint}. Statistics computed from historical behavior use only information available before the current time step. Such candidate-conditioned modeling is widely used in behavior-aware
ranking models \citep{zhou2019deep}.
\item $x^{A}_{u,t}$ denotes the Action information, namely the user feedback generated after item exposure or interaction, including clicks, dwell time, collections, purchases, and ratings \citep{hu2008collaborative,rendle2012bpr}. Historical interactions use the observed Action, while the ground-truth Action at the target position is excluded from the input and used only to construct the training supervision or evaluation label.
\end{itemize}
The four types of fields therefore follow this lifecycle order:
$\mathrm{Request}
\rightarrow
\mathrm{Item}
\rightarrow
\mathrm{Context}
\rightarrow
\mathrm{Action}.$
This order reflects how information becomes available during a recommendation interaction: Request specifies the decision conditions, Item identifies the interaction object, Context describes the candidate under the current scenario, and Action records the final user response. This decomposition provides multiple semantically distinct state positions within the same interaction, allowing retrieval and ranking to read historical representations corresponding to their respective prediction stages.

For any role $q$ in the role set $\mathcal{Q}=\{R,I,C,A\}$, we further divide its input fields into a set of sparse fields $\mathcal{S}_{q}$ and a set of dense fields $\mathcal{D}_{q}$. Sparse fields are encoded using independent embedding tables, while dense fields are first normalized and then transformed by a dense-feature encoder. The two types of representations are aggregated separately, concatenated, and projected into a shared $d$-dimensional hidden space \citep{rendle2010factorization,naumov2019deep}:
\begin{equation}
\begin{aligned}
\mathbf{s}^{q}_{u,t}
&=
\operatorname{Agg}_{\mathrm{sparse}}
\left(
\left\{
\mathbf{E}_{q,k}
\left[
x^{q,k}_{u,t}
\right]
\right\}_{k\in\mathcal{S}_{q}}
\right),
\\
\mathbf{d}^{q}_{u,t}
&=
\operatorname{Enc}_{\mathrm{dense},q}
\left(
\operatorname{Concat}_{\ell\in\mathcal{D}_{q}}
\operatorname{Norm}
\left(
x^{q,\ell}_{u,t}
\right)
\right),
\\
\mathbf{z}^{q}_{u,t}
&=
\operatorname{Proj}_{q}
\left(
\left[
\mathbf{s}^{q}_{u,t};
\mathbf{d}^{q}_{u,t}
\right]
\right)
\in\mathbb{R}^{d},
\qquad
q\in\mathcal{Q}.
\end{aligned}
\label{eq:role_feature_encoding}
\end{equation}
Here, $\mathbf{E}_{q,k}$ denotes the embedding table for the $k$-th sparse field of role $q$, $\operatorname{Agg}_{\mathrm{sparse}}(\cdot)$ denotes the aggregation of sparse-field representations, $\operatorname{Norm}(\cdot)$ denotes dense-feature normalization, $\operatorname{Enc}_{\mathrm{dense},q}(\cdot)$ denotes the dense-feature encoder, and $\operatorname{Proj}_{q}(\cdot)$ projects the concatenated representation to the fixed hidden dimension $d$. Consequently, different roles can use role-specific input fields while producing lifecycle tokens in the same hidden space.

LIFT then expands each interaction into four tokens according to the lifecycle order and concatenates all historical interactions into a unified sequence:
\begin{equation}
\begin{aligned}
\mathbf{Z}_{u,\leq T}
&=
\operatorname{Serialize}
\left(
e_{u,1:T}
\right)
\\
&=
\left[
\mathbf{z}^{R}_{u,1},
\mathbf{z}^{I}_{u,1},
\mathbf{z}^{C}_{u,1},
\mathbf{z}^{A}_{u,1},
\ldots,
\mathbf{z}^{R}_{u,T},
\mathbf{z}^{I}_{u,T},
\mathbf{z}^{C}_{u,T},
\mathbf{z}^{A}_{u,T}
\right]
\in\mathbb{R}^{4T\times d}.
\end{aligned}
\label{eq:lifecycle_serialization}
\end{equation}
The lifecycle order also determines the availability of information within an interaction. Let $R \prec I \prec C \prec A$ denote the precedence relation among the four roles. Under causal modeling, the hidden state at each position can depend only on the current token and preceding tokens in the serialized sequence. Roles that occur later in the lifecycle do not contribute to its computation. Consequently, a Request state primarily reflects the historical information available when the request is formed, while a Context state can additionally use the Item and Context information of the current interaction. At a prediction position, the ground-truth Action is excluded from the input and is used only to construct the training supervision. The detailed causal-attention constraints and lifecycle-role-specific state transformations are presented in Section~\ref{sec:role_conditioned_attention}.

The lifecycle states provide natural prediction boundaries for the two
stages. Retrieval is performed before a candidate item is selected, so its
representation should depend on the historical sequence and the current
Request, but not on the Item, Context, or Action of the current interaction.
The Request state therefore forms the retrieval readout. Ranking is applied
after a candidate item has been introduced and its candidate-dependent
features are available, while the user Action remains unknown. The Context
state consequently forms the ranking readout. These two states are produced
by the same causal sequence model but correspond to different information
sets, allowing shared historical modeling without collapsing the stage
boundary between retrieval and ranking.

Through this representation, each interaction is organized as four tokens with an explicit lifecycle order. The resulting sequence preserves both cross-interaction historical dependencies and information availability within each interaction, providing retrieval and ranking with a shared input space and semantically distinct state readout positions.
\subsection{Role-Conditioned Attention and Pre-LN Bias}
\label{sec:role_conditioned_attention}

Given the lifecycle sequence introduced in Section~\ref{sec:lifecycle_representation}, let $L=4T$ denote the number of lifecycle tokens. Let $\mathcal{Q}=\{R,I,C,A\}$ be the set of lifecycle roles and $\mathcal{Q}_{\mathrm{ord}}=(R,I,C,A)$ their fixed order. For token position $i\in\{1,\ldots,L\}$, its lifecycle role and interaction position are defined as
\begin{equation}
\rho(i)
=
\left(\mathcal{Q}_{\mathrm{ord}}\right)_{1+((i-1)\bmod 4)},
\qquad
\tau(i)
=
\left\lfloor\frac{i-1}{4}\right\rfloor+1.
\end{equation}
To simplify notation, we omit the user and sequence subscripts and write
$\mathbf{Z}=\mathbf{Z}_{u,\leq T}$ and
$\mathbf{H}^{(B)}=\mathbf{H}_{u,\leq T}$.
Let $b\in\{1,\ldots,B\}$ index the Decoder layers, and denote the input token at position $i$ by $\mathbf{z}_i=\mathbf{z}^{\rho(i)}_{u,\tau(i)}$.

The LIFT Decoder takes the input tokens as its zeroth-layer hidden states and produces position-aligned hidden states after $B$ Decoder layers:
\begin{equation}
\begin{aligned}
\mathbf{H}^{(0)}
&=
\mathbf{Z}
=
[\mathbf{z}_1,\ldots,\mathbf{z}_L],
\\
\mathbf{H}^{(B)}
&=
F_{\theta}(\mathbf{Z})
=
\left[
\mathbf{h}^{(B)}_1,
\mathbf{h}^{(B)}_2,
\ldots,
\mathbf{h}^{(B)}_L
\right].
\end{aligned}
\label{eq:decoder_output}
\end{equation}

To make the parameter-sharing boundary explicit, we partition the Decoder parameters into those shared across lifecycle roles and those indexed by lifecycle role:
\begin{equation}
\begin{aligned}
\theta_{\mathrm{shared}}
&=
\left\{
\mathbf{W}^{(b)}_{Q},
\mathbf{W}^{(b)}_{K},
\theta^{(b)}_{\mathrm{norm}},
\theta^{(b)}_{\mathrm{FFN}}
\;\middle|\;
b=1,\ldots,B
\right\},
\\
\theta_{\mathrm{role}}
&=
\left\{
\mathbf{r}^{(b)}_{q},
\mathbf{W}^{(b,q)}_{V},
\mathbf{W}^{(b,q)}_{O},
\mathbf{b}^{(b)}_{q}
\;\middle|\;
b=1,\ldots,B,\ q\in\mathcal{Q}
\right\},
\\
\theta
&=
\theta_{\mathrm{shared}}
\cup
\theta_{\mathrm{role}}.
\end{aligned}
\label{eq:decoder_parameter_partition}
\end{equation}
Here, $\theta^{(b)}_{\mathrm{norm}}$ collects the normalization parameters in layer $b$, and $\theta^{(b)}_{\mathrm{FFN}}$ denotes the parameters of its feed-forward sublayer. The vectors $\mathbf{r}^{(b)}_q$ and $\mathbf{b}^{(b)}_q$ are the learnable role embedding and additive bias associated with role $q$, respectively. Task-specific prediction heads are defined separately in Section~\ref{sec:overview}.

The lifecycle roles differ primarily in the information they carry and the
states they produce, while their dependencies are modeled in a common
sequence space. We therefore share Query and Key projections for dependency
matching, and condition Value and Output projections on the source and
receiving roles, respectively.

At layer $b$, Role-Conditioned Attention models dependencies across interactions and lifecycle roles. Given the input hidden state $\mathbf{h}^{(b-1)}_i$, Query, Key, and Value are constructed through the following pre-normalization paths:
\begin{equation}
\begin{aligned}
\widetilde{\mathbf{h}}^{V,(b)}_i
&=
\operatorname{LN}^{(b)}_{\mathrm{attn}}
\left(
\mathbf{h}^{(b-1)}_i
\right),
\\
\widetilde{\mathbf{h}}^{QK,(b)}_i
&=
\operatorname{LN}^{(b)}_{\mathrm{attn}}
\left(
\mathbf{h}^{(b-1)}_i
+
\mathbf{r}^{(b)}_{\rho(i)}
\right),
\\
\mathbf{q}^{(b)}_i
&=
\widetilde{\mathbf{h}}^{QK,(b)}_i
\mathbf{W}^{(b)}_{Q},
\qquad
\mathbf{k}^{(b)}_i
=
\widetilde{\mathbf{h}}^{QK,(b)}_i
\mathbf{W}^{(b)}_{K},
\\
\mathbf{v}^{(b)}_i
&=
\widetilde{\mathbf{h}}^{V,(b)}_i
\mathbf{W}^{(b,\rho(i))}_{V}.
\end{aligned}
\label{eq:role_conditioned_qkv}
\end{equation}
The role vector $\mathbf{r}^{(b)}_{\rho(i)}$ is a learnable identity vector associated with the lifecycle role at position $i$. Query and Key are computed from the normalized state augmented with role information, whereas Value is computed from the normalized state without the role embedding. The Value projection is selected according to the role of the source token.

Attention scores are computed using an event-position-aware Query--Key compatibility function:
\begin{equation}
\ell^{(b)}_{ij}
=
\frac{
\Psi_{\mathrm{evt}}
\left(
\mathbf{q}^{(b)}_i,
\mathbf{k}^{(b)}_j;
\tau(i),\tau(j)
\right)
}{
\sqrt{d_h}
}
+
M_{ij},
\qquad
M_{ij}
=
\begin{cases}
0, & j\leq i,\\
-\infty, & j>i.
\end{cases}
\label{eq:event_position_attention_score}
\end{equation}
Here,
$\Psi_{\mathrm{evt}}(\mathbf{q}_i,\mathbf{k}_j;\tau(i),\tau(j))
=
(\mathbf{R}_{\tau(i)}\mathbf{q}_i)^{\top}
(\mathbf{R}_{\tau(j)}\mathbf{k}_j)$
denotes RoPE-based Query--Key compatibility
\citep{su2024roformer}, where
$\mathbf{R}_{\tau}$ is the rotary transformation at interaction position
$\tau$. Thus, lifecycle tokens from the same interaction share the same
rotary position, and $M_{ij}$ is the causal mask.

Let
$\boldsymbol{\ell}^{(b)}_{i,1:L}
=
(\ell^{(b)}_{i1},\ldots,\ell^{(b)}_{iL})$
denote the attention-score vector from position $i$ to the full sequence. The attention output and residual connection are defined as
\begin{equation}
\mathbf{u}^{(b)}_i
=
\mathbf{h}^{(b-1)}_i
+
\left[
\sum_{j=1}^{L}
\left[
\operatorname{softmax}
\left(
\boldsymbol{\ell}^{(b)}_{i,1:L}
\right)
\right]_j
\mathbf{v}^{(b)}_j
\right]
\mathbf{W}^{(b,\rho(i))}_{O}.
\label{eq:role_conditioned_attention_output}
\end{equation}
The output projection $\mathbf{W}^{(b,\rho(i))}_{O}$ is selected according to the lifecycle role of the receiving position. Thus, the Value projection depends on the source role, whereas the Output projection depends on the receiving role. Role-Conditioned Attention models dependencies across interactions and lifecycle roles in a shared Query--Key matching space while preserving role-dependent information transformations.

Following the Pre-LN Transformer organization
\citep{xiong2020layer}, LIFT adds a role-specific bias on the
pre-normalization path to the feed-forward sublayer. Let $\mathbf{u}^{(b)}_i$ denote the output of the attention sublayer. The FFN input and the resulting hidden state are
\begin{equation}
\begin{aligned}
\widetilde{\mathbf{u}}^{(b)}_i
&=
\operatorname{LN}^{(b)}_{\mathrm{FFN}}
\left(
\mathbf{u}^{(b)}_i
+
\mathbf{b}^{(b)}_{\rho(i)}
\right),
\\
\mathbf{h}^{(b)}_i
&=
\mathbf{u}^{(b)}_i
+
\operatorname{FFN}^{(b)}
\left(
\widetilde{\mathbf{u}}^{(b)}_i
\right).
\end{aligned}
\label{eq:preln_bias_ffn}
\end{equation}
The final Request and Context states are extracted by the task-specific readout operators defined in Section~\ref{sec:overview}
 and passed to the retrieval and ranking heads, respectively.

\section{Experiments}

\subsection{Experimental Setup}
\label{sec:experimental_setup}

\begin{wraptable}{r}{\sideTableWidth}
\vspace{-4\baselineskip}
\centering
\compacttable
\caption{Dataset statistics for ML-20M and Taobao.}
\label{tab:dataset_stats}
\resizebox{\linewidth}{!}{%
\begin{tabular}{lrrrrr}
\toprule
& \multicolumn{2}{c}{Dataset size}
& \multicolumn{3}{c}{Interactions} \\
\cmidrule(lr){2-3}
\cmidrule(lr){4-6}
Dataset & Users & Items & Train & Validation & Test \\
\midrule
ML-20M
& 138,493 & 27,278
& 19,153,489 & 562,888 & 283,886 \\
Taobao
& 961,904 & 1,518,629
& 68,556,084 & 12,607,812 & 12,423,797 \\
\bottomrule
\end{tabular}%
}
\vspace{-2\baselineskip}

\end{wraptable}
\paragraph{Datasets.}
We evaluate LIFT on ML-20M \citep{harper2015movielens} and
Taobao \citep{alibaba2017taobao}. Table~\ref{tab:dataset_stats} summarizes the dataset statistics. The catalog size excludes the padding ID.

Both datasets are split chronologically. For ML-20M, training interactions span 1995-01-09 to 2013-12-31, validation covers 2014-01-01 to 2014-12-31, and test covers 2015-01-01 to 2015-03-31. For Taobao, training covers 2017-11-25 to 2017-12-01, validation is 2017-12-02, and test is 2017-12-03. We retain Taobao items with at least 5 training-period interactions and users with at least 10 retained interactions, including at least 1 training interaction \citep{wang2018billion}.
\paragraph{Tasks and evaluation protocol.}
The retrieval task predicts the next interacted item and is trained with
an InfoNCE objective \citep{oord2018representation} over in-batch target items, with LogQ correction
\citep{yi2019sampling}. At evaluation, we rank items from the available catalog and filter items previously interacted with by the user. We report Recall@100 and NDCG@100. The ranking task predicts feedback for a target interaction and is trained with binary cross-entropy. For ML-20M, ratings greater than 4 are positive labels. For Taobao, purchases are positive, and the other behavior types are negative. We report AUC \citep{fawcett2006introduction} and GAUC \citep{zhou2018deep}. For joint recommendation, we define the Joint Score as
\begin{equation}
\mathrm{JointScore}
=
\sqrt{\mathrm{NDCG@100}\cdot\mathrm{GAUC}}.
\end{equation}
The geometric mean treats retrieval and ranking symmetrically and penalizes
solutions that perform poorly on either stage.
\paragraph{Baselines.}
Retrieval baselines include YouTubeDNN
\citep{covington2016deep}, GRU4Rec
\citep{hidasi2015session}, SASRec
\citep{kang2018self}, HSTU
\citep{zhai2024actions}, and UniPinRec
\citep{li2026unipinrec}. Ranking baselines include DeepFM
\citep{guo2017deepfm}, DIN
\citep{zhou2018deep}, DCNv2
\citep{wang2021dcn}, RankMixer
\citep{zhu2025rankmixer}, HSTU, and UniPinRec. Joint baselines include
SharedBottom \citep{caruana1997multitask}, MMoE
\citep{ma2018modeling}, PLE
\citep{tang2020progressive}, HSTU, and UniPinRec. OneRec and OnePiece represent generative and latent-reasoning formulations,
respectively, and are discussed in Related Work. Our direct comparison
focuses on methods evaluated under the common catalog-based
retrieval--ranking protocol.

Within each task, all methods use the same chronological splits, candidate
construction, supervision, and evaluation protocol. Sequential models use
the same maximum interaction history, with hidden dimension and depth
aligned where applicable, while retaining their native input interfaces
and prediction heads. Checkpoints are selected by validation NDCG@100 for
retrieval, AUC for ranking, and Joint Score for joint training; the test set
is used only for final evaluation. Unless otherwise specified, LIFT uses 100 historical interactions,
three Transformer layers, hidden dimension 256, and eight attention heads.
All reported quality metrics are averaged over random seeds 2026, 2027,
and 2028.

\paragraph{Overall Results.}
Table~\ref{tab:en_joint} shows that LIFT's Joint Score reaches 0.2724 on ML-20M,
improving over HSTU by 4.9\%, and 0.2239 on Taobao, improving over
UniPinRec by 3.6\%. Dedicated retrieval and ranking results are reported in
Appendix~\ref{app:dedicated_results}.

\begin{table*}[t]
\centering
\caption{Joint Retrieval--Ranking results on ML-20M and Taobao. Joint Score is defined as
$\sqrt{\mathrm{NDCG@100}\cdot\mathrm{GAUC}}$.}
\label{tab:en_joint}
\compacttable
\begin{adjustbox}{max width=\textwidth}
\begin{tabular}{lrrrrr@{\qquad}rrrrr}
\toprule
& \multicolumn{5}{c}{ML-20M}
& \multicolumn{5}{c}{Taobao} \\
\cmidrule(lr){2-6}
\cmidrule(lr){7-11}
Method
& R@100 & N@100 & AUC & GAUC & Joint
& R@100 & N@100 & AUC & GAUC & Joint \\
\midrule
SharedBottom
& 0.2603 & 0.0681 & 0.7748 & 0.7058 & 0.2193
& 0.1518 & 0.0454 & 0.8950 & 0.8575 & 0.1972 \\
MMoE
& 0.2702 & 0.0724 & 0.7928 & 0.7105 & 0.2268
& 0.1552 & 0.0467 & 0.9001 & 0.8588 & 0.2002 \\
PLE
& 0.2004 & 0.0483 & 0.7649 & 0.6451 & 0.1764
& 0.1521 & 0.0455 & 0.8962 & 0.8556 & 0.1973 \\
HSTU
& 0.3157 & 0.0910 & 0.8368 & 0.7410 & 0.2597
& 0.1716 & 0.0526 & 0.9099 & 0.8570 & 0.2123 \\
UniPinRec
& 0.3152 & 0.0894 & 0.8442 & 0.7460 & 0.2582
& 0.1815 & 0.0551 & 0.9036 & 0.8482 & 0.2162 \\
LIFT
& \textbf{0.3361} & \textbf{0.0981} & \textbf{0.8496}
& \textbf{0.7565} & \textbf{0.2724}
& \textbf{0.1885} & \textbf{0.0582} & \textbf{0.9127}
& \textbf{0.8612} & \textbf{0.2239} \\
\bottomrule
\end{tabular}
\end{adjustbox}
\vspace{-2\baselineskip}
\end{table*}

\subsection{Retrieval--Ranking Pareto Frontier}

\begin{figure}[h]
    \vspace{-2\baselineskip}
    \centering
    \includegraphics[width=\linewidth]{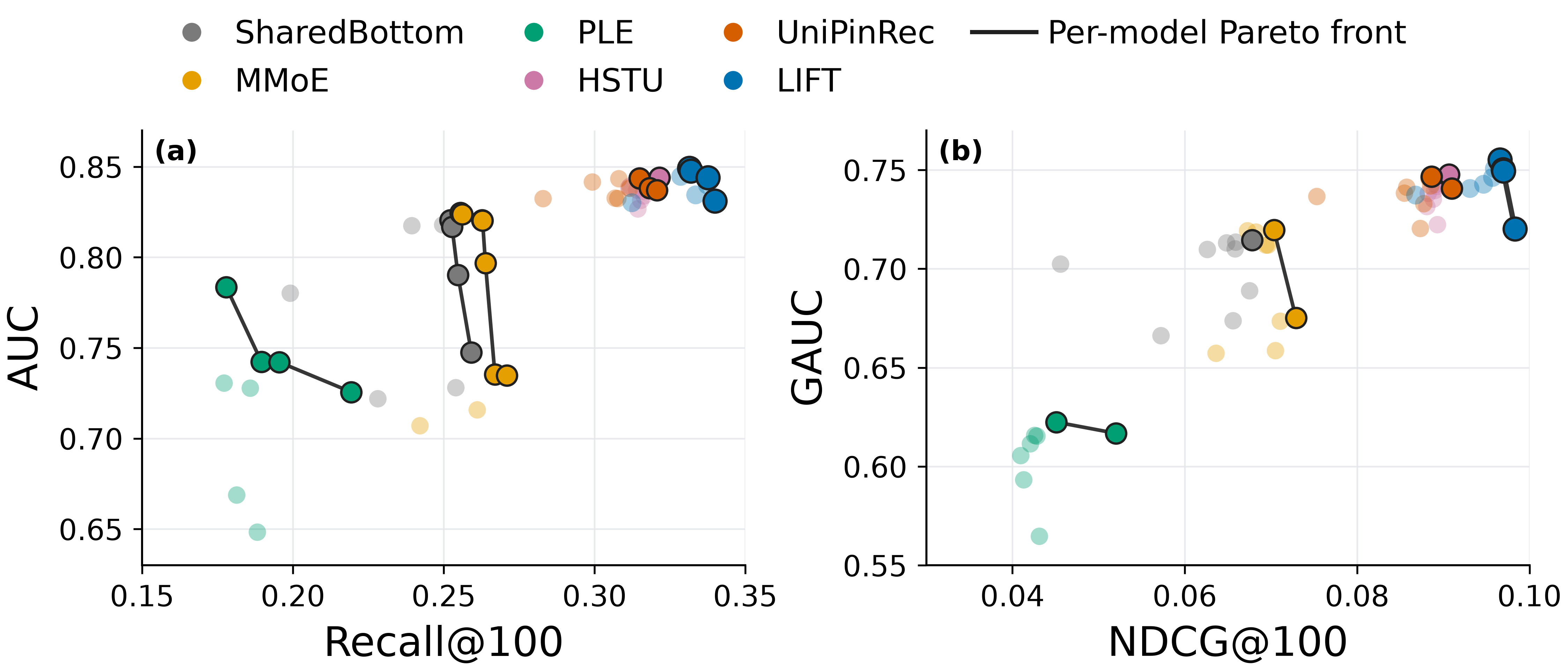}
    \vspace{-2\baselineskip}
    \caption{Retrieval--ranking trade-offs on ML-20M. (a) Recall@100 versus AUC; (b) NDCG@100 versus GAUC. Colors denote different methods, and black line segments connect each method's non-dominated configurations. Higher values are better on both axes.}
    \label{fig:pareto}
    \vspace{-0\baselineskip}
\end{figure}

To examine retrieval--ranking trade-offs, for each joint model, we vary $\lambda_{\mathrm{ret}}$ over 
\{0.0078125, 0.015625, 0.03125, 0.0625, 0.125, 0.25, 0.5, 1, 2\}
while fixing $\lambda_{\mathrm{rank}}=1$ on ML-20M. Each point in
Figure~\ref{fig:pareto} reports the test performance of the checkpoint
selected by validation Joint Score, and the connected non-dominated points
form each method's empirical Pareto frontier.

Increasing $\lambda_{\mathrm{ret}}$ shifts LIFT toward retrieval-oriented
operating points, while moderate weights preserve stronger ranking quality.
Across the evaluated loss weights, LIFT configurations form the global
empirical Pareto frontier in both Recall@100--AUC and NDCG@100--GAUC
spaces, indicating stable retrieval--ranking trade-offs across different
operating points.

\subsection{Ablation Studies}
\par
\WFclear
\begin{wraptable}{r}{\sideTableWidth}
\vspace{-4\baselineskip}
\centering
\compacttable
\caption{Component ablations on ML-20M. Higher values are better for all metrics.}
\label{tab:component_ablation}
\resizebox{\linewidth}{!}{%
\begin{tabular}{lrrrrr}
\toprule
Variant & R@100 & N@100 & AUC & GAUC & Joint Score \\
\midrule
Basic Transformer
& 0.3255 & 0.0941 & 0.8464 & 0.7535 & 0.2663 \\
+ Pre-LN Bias
& 0.3330 & 0.0963 & 0.8463 & 0.7541 & 0.2695 \\
+ Role-Cond. Attn.
& 0.3292 & 0.0947 & 0.8480 & 0.7537 & 0.2672 \\
\textbf{LIFT}
& \textbf{0.3361} & \textbf{0.0981} & \textbf{0.8496} & \textbf{0.7565} & \textbf{0.2724} \\
\bottomrule
\end{tabular}%
}
\vspace{-2\baselineskip}
\end{wraptable}

\paragraph{Component Ablations.}
Table~\ref{tab:component_ablation} uses a $2\times2$ design to isolate
Role-Conditioned Attention and Pre-LN Bias while keeping the lifecycle
sequence, task readouts, heads, and training protocol fixed.

Removing Pre-LN Bias reduces Joint Score from 0.2724 to 0.2672, removing
Role-Conditioned Attention reduces it to 0.2695, and removing both yields
0.2663. The full model performs best on all reported metrics, supporting
complementary contributions from the two components.

\paragraph{Lifecycle Sequence-Structure Ablation.}
Table~\ref{tab:sequence_ablation} compares alternative lifecycle groupings
and task readouts under the same encoder and training protocol. Multiple
letters within one pair of brackets denote a merged token, while
\texttt{[]} denotes an explicit query token. For \texttt{[RICA]}, retrieval reads the preceding interaction state,
while ranking reads the current event state with the target Action masked.

\par
\WFclear
\begin{table}[h]
\vspace{-1\baselineskip}
\centering
\compacttable
\caption{Lifecycle sequence-construction ablations on ML-20M. Higher values are better for all metrics.}
\label{tab:sequence_ablation}
\resizebox{\linewidth}{!}{%
\begin{tabular}{lllrrrrr}
\toprule
Sequence & Ret. Readout & Rank. Readout
& R@100 & N@100 & AUC & GAUC & Joint Score \\
\midrule
\texttt{[RICA]}
& $\texttt{[RICA]}_{t-1}$
& $\texttt{[RICA]}_{t}$
& 0.3152 & 0.0894 & 0.8442 & 0.7460 & 0.2582 \\

\texttt{[RIC][A]}
& $\texttt{[A]}_{t-1}$
& \texttt{[RIC]}
& 0.3195 & 0.0921 & 0.8457 & 0.7533 & 0.2634 \\

$\texttt{[]}_{I}\texttt{[RIC][A]}$
& $\texttt{[]}_{I}$
& \texttt{[RIC]}
& 0.3226 & 0.0931 & 0.8464 & 0.7532 & 0.2647 \\

\texttt{[R][IC][A]}
& \texttt{[R]}
& \texttt{[IC]}
& 0.3325 & 0.0952 & 0.8419 & 0.7390 & 0.2652 \\

$\texttt{[]}_{I}\texttt{[RIC]}\texttt{[]}_{A}\texttt{[A]}$
& $\texttt{[]}_{I}$
& $\texttt{[]}_{A}$
& 0.3219 & 0.0922 & 0.8488 & 0.7553 & 0.2639 \\

\textbf{\texttt{[R][I][C][A]}}
& \textbf{\texttt{[R]}}
& \textbf{\texttt{[C]}}
& \textbf{0.3361}
& \textbf{0.0981}
& \textbf{0.8496}
& \textbf{0.7565}
& \textbf{0.2724} \\
\bottomrule
\end{tabular}%
}
\vspace{-1\baselineskip}
\end{table}

The fully merged \texttt{[RICA]} representation obtains a Joint Score of
0.2582, whereas the fully factorized \texttt{[R][I][C][A]} sequence reaches
0.2724. The alternative four-token layout $\texttt{[]}_{I}\texttt{[RIC]}\texttt{[]}_{A}\texttt{[A]}$ reaches
0.2639. The gap between the two four-token layouts suggests that aligning
token boundaries with information-availability stages is important for
constructing effective retrieval and ranking states.

\subsection{Capacity and Sequence Scaling Analysis}
\label{sec:scaling_analysis}

\begin{figure}[h]
    \vspace{-1\baselineskip}
    \centering
    \includegraphics[width=\linewidth]{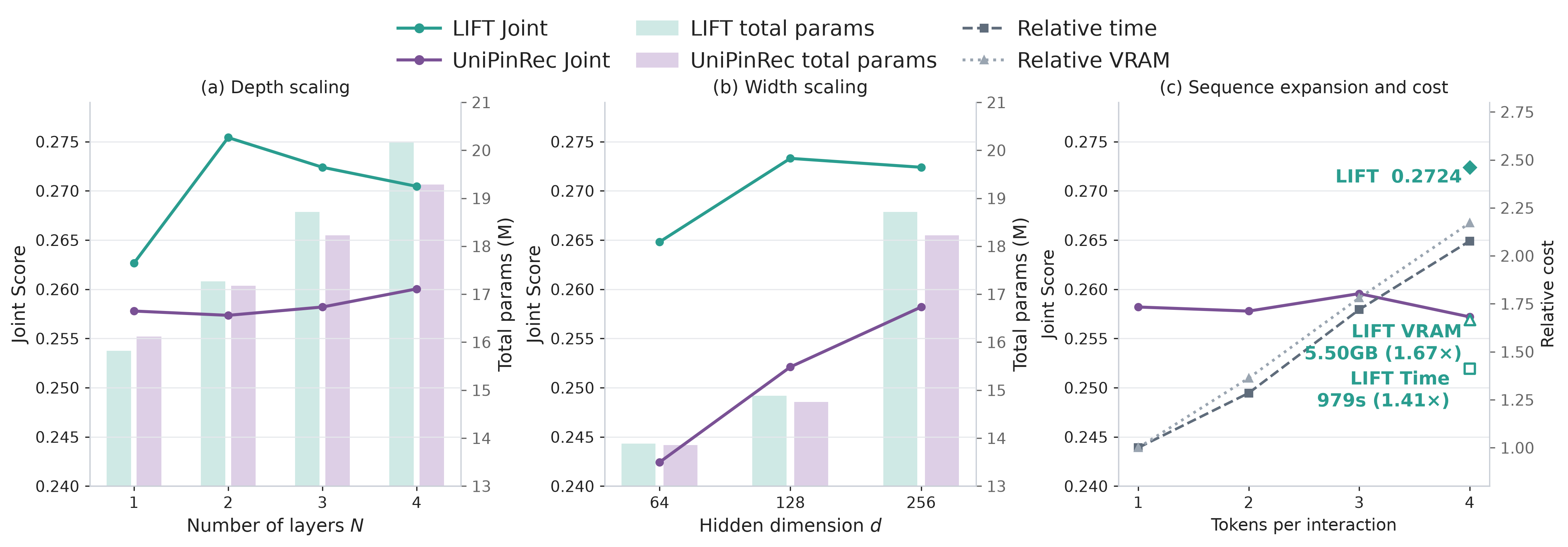}
    \vspace{-2\baselineskip}
    \caption{
        Capacity and sequence scaling on ML-20M.
        (a) Depth scaling.
        (b) Width scaling.
        (c) Sequence expansion and computational cost.
        }
    \label{fig:scaling}
    \vspace{-1\baselineskip}
\end{figure}

Figure~\ref{fig:scaling} compares LIFT and UniPinRec under
depth, width, and sequence expansion.

\paragraph{Capacity scaling.}
LIFT achieves a higher Joint Score than UniPinRec at every evaluated depth
and hidden dimension. Notably, LIFT with $d=64$ reaches 0.2648 using
13.88M parameters, exceeding UniPinRec with $d=256$, which reaches 0.2582
with 18.23M parameters.

\paragraph{Sequence expansion.}
Adding blank tokens to UniPinRec leaves Joint Score nearly unchanged
(0.2572--0.2596), whereas LIFT reaches 0.2724 with four semantic lifecycle
tokens. At the matched four-token sequence length and the same encoder
depth and hidden dimension, LIFT completes the 24-epoch training run in
979.0 seconds versus 1440.8 seconds for UniPinRec with three blank tokens,
a 32.1\% reduction, while peak memory decreases from 7.16 GB to 5.50 GB.

\section{Conclusion}

We presented LIFT, a unified retrieval--ranking model built around
lifecycle-aware interaction factorization. LIFT decomposes each interaction
into Request, Item, Context, and Action states according to information
availability, allowing retrieval and ranking to share causal history modeling
while reading representations aligned with their respective prediction
stages. Role-conditioned attention and a lightweight Pre-LN Bias instantiate
this factorization within a shared Transformer backbone.
Experiments on ML-20M and Taobao show consistent improvements in joint
retrieval--ranking quality, while loss-weight sweeps demonstrate favorable
trade-offs between the two tasks. Lifecycle sequence ablations and scaling
analyses further characterize the effects of sequence construction and
model capacity.

\section*{AI Use Statement}

Generative AI tools were used during manuscript preparation for language
polishing, consistency checking, code assistance, and feedback on experimental
presentation. The authors independently verified all technical
content, implementations, experimental results, analyses, and conclusions,
and take full responsibility for the content of the paper.

\bibliography{iclr2027_conference}
\bibliographystyle{iclr2027_conference}
\newpage

\appendix

\section{Dataset Feature Assignment}
\label{app:dataset_features}

Table~\ref{tab:dataset_features} summarizes the lifecycle-role assignment
used in our experiments. Derived historical statistics are computed causally
using only information available before the current interaction. The target
Action is excluded from the input at the prediction position and is used only
to construct supervision.

\begin{table*}[h]
\centering
\caption{Lifecycle feature assignment used in ML-20M and Taobao.}
\label{tab:dataset_features}
\compacttable
\begin{tabular}{lllll}
\toprule
Dataset & Request & Item & Context & Action \\
\midrule
ML-20M
& weekend
& item ID
& item count (log)
& rating \\
& user activity
& release year
& average rating
& positive label \\
& time since last
& genres
& & \\
\midrule
Taobao
& user activity (log2)
& item ID
& item count (log)
& behavior type \\
& time gap
& category
& user--item count
& purchase label \\
\bottomrule
\end{tabular}
\end{table*}

For ML-20M, the positive ranking label is defined as a rating greater than
4. For Taobao, purchase is treated as positive feedback and the other
behavior types as negative feedback. Item-count, user-activity,
user--item-count, average-rating, and time-gap features are constructed
from historical observations only. Sequence-based models do not expose
the current target Action to the encoder.

\section{Hyperparameter Search Space}
\label{app:hyperparameters}

Table~\ref{tab:hyperparameter_space} summarizes the hyperparameter values
considered during model selection. We use coordinate search and select
configurations according to validation performance.

\begin{table}[h]
\centering
\caption{Hyperparameter search spaces used in the experiments.}
\label{tab:hyperparameter_space}
\compacttable
\begin{tabular}{lll}
\toprule
Task & Hyperparameter & Search space \\
\midrule
Retrieval
& Learning rate
& $\{1\mathrm{e}{-4},\,3\mathrm{e}{-4},\,1\mathrm{e}{-3},\,2\mathrm{e}{-3},\,3\mathrm{e}{-3}\}$ \\
& Dropout
& $\{0.0,\,0.1,\,0.2,\,0.3,\,0.4,\,0.5\}$ \\
& Weight decay
& $\{0.0,\,0.001,\,0.01\}$ \\
& Temperature
& $\{0.05,\,0.07,\,0.10\}$ \\
& LogQ $\alpha$
& $\{0.50,\,0.75,\,1.00\}$ \\
\midrule
Ranking
& Learning rate
& $\{1\mathrm{e}{-4},\,3\mathrm{e}{-4},\,1\mathrm{e}{-3},\,2\mathrm{e}{-3},\,3\mathrm{e}{-3}\}$ \\
& Dropout
& $\{0.0,\,0.1,\,0.2,\,0.3,\,0.4,\,0.5\}$ \\
& Weight decay
& $\{0.0,\,0.001,\,0.01\}$ \\
\midrule
Joint
& Learning rate
& $\{1\mathrm{e}{-4},\,3\mathrm{e}{-4},\,1\mathrm{e}{-3},\,2\mathrm{e}{-3},\,3\mathrm{e}{-3}\}$ \\
& Dropout
& $\{0.0,\,0.1,\,0.2,\,0.3,\,0.4,\,0.5\}$ \\
& Weight decay
& $\{0.0,\,0.001,\,0.01\}$ \\
& Temperature
& $\{0.05,\,0.07,\,0.10\}$ \\
& LogQ $\alpha$
& $\{0.50,\,0.75,\,1.00\}$ \\
\bottomrule
\end{tabular}
\end{table}
\section{Dedicated Single-Task Results}
\label{app:dedicated_results}

We additionally evaluate all methods under dedicated retrieval and ranking
training. Table~\ref{tab:dedicated_results} reports the results on ML-20M
and Taobao.

\begin{table}[h]
\centering
\compacttable
\caption{Dedicated retrieval and ranking results on ML-20M and Taobao.}
\label{tab:dedicated_results}

\begin{minipage}[t]{0.48\textwidth}
\centering
\textbf{(a) Retrieval}
\vspace{0.3em}

\begin{tabular*}{\linewidth}{@{\extracolsep{\fill}}lrrrr@{}}
\toprule
& \multicolumn{2}{c}{ML-20M} & \multicolumn{2}{c}{Taobao} \\
\cmidrule(lr){2-3}
\cmidrule(lr){4-5}
Method & R@100 & N@100 & R@100 & N@100 \\
\midrule
YouTubeDNN & 0.2215 & 0.0554 & 0.0685 & 0.0191 \\
GRU4Rec    & 0.3142 & 0.0909 & 0.1570 & 0.0466 \\
SASRec     & 0.3110 & 0.0902 & 0.1609 & 0.0479 \\
HSTU       & 0.3142 & 0.0889 & 0.1684 & 0.0511 \\
UniPinRec  & 0.3186 & 0.0916 & 0.1694 & 0.0510 \\
LIFT       & \textbf{0.3352} & \textbf{0.0961}
           & \textbf{0.1790} & \textbf{0.0544} \\
\bottomrule
\end{tabular*}
\end{minipage}
\hfill
\begin{minipage}[t]{0.48\textwidth}
\centering
\textbf{(b) Ranking}
\vspace{0.3em}

\begin{tabular*}{\linewidth}{@{\extracolsep{\fill}}lrrrr@{}}
\toprule
& \multicolumn{2}{c}{ML-20M} & \multicolumn{2}{c}{Taobao} \\
\cmidrule(lr){2-3}
\cmidrule(lr){4-5}
Method & AUC & GAUC & AUC & GAUC \\
\midrule
DeepFM    & 0.7116 & 0.6851 & 0.8907 & \textbf{0.8692} \\
DIN       & 0.7200 & 0.7061 & 0.8894 & 0.8664 \\
DCNv2     & 0.7087 & 0.6937 & 0.8838 & 0.8645 \\
RankMixer & 0.8170 & 0.7088 & 0.9006 & 0.8361 \\
HSTU      & 0.8480 & 0.7530 & 0.9070 & 0.8588 \\
UniPinRec & 0.8493 & 0.7545 & 0.9086 & 0.8493 \\
LIFT      & \textbf{0.8509} & \textbf{0.7569}
           & \textbf{0.9129} & 0.8587 \\
\bottomrule
\end{tabular*}
\end{minipage}

\end{table}

LIFT achieves the strongest retrieval results on both datasets and the
highest ranking AUC on both. It also obtains the highest GAUC on ML-20M;
on Taobao, DeepFM achieves the highest GAUC, while LIFT retains the
highest AUC. These results show that lifecycle factorization remains
effective when retrieval and ranking are trained independently.

\section{Causal Mask with Multiple Exposures}
\label{app:multi_exposure_mask}

In practical recommendation logs, a single request event may contain
multiple exposure samples corresponding to different candidate items \citep{wu2020mind, ie2019slateq}.
These candidates are generated under the same request and have no temporal
precedence relation with each other. Directly applying a standard
lower-triangular mask would therefore introduce an artificial ordering
between co-occurring candidates. We extend the causal mask to preserve
temporal causality across events while isolating different exposures within
the same event. The formulation in the main text corresponds to the single-exposure case.

\begin{figure}[h]
    \centering
    \includegraphics[width=\linewidth]{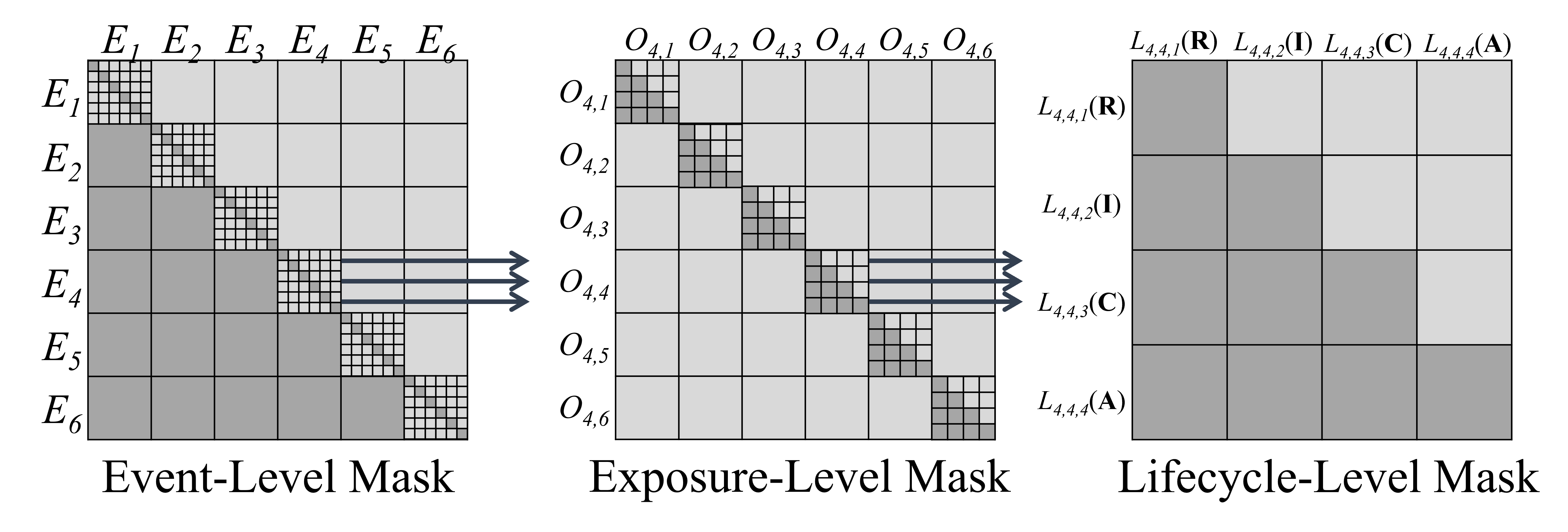}
    \caption{Causal mask construction for events containing multiple
    exposure samples. Visibility is determined by event-level temporal
    causality, exposure-level isolation, and lifecycle-level ordering.}
    \label{fig:multi_exposure_mask}
\end{figure}

Let $q$ and $k$ denote the query and key token positions, respectively.
Each token position is associated with three indices:
\begin{equation}
(e_q,o_q,c_q),
\qquad
(e_k,o_k,c_k),
\end{equation}
where $e$ denotes the event index, $o$ denotes the exposure index within
an event, and $c$ denotes the lifecycle-stage index. We assign
\begin{equation}
c(R)=1,\qquad
c(I)=2,\qquad
c(C)=3,\qquad
c(A)=4,
\end{equation}
following the lifecycle order
\[
\mathrm{Request}
\rightarrow
\mathrm{Item}
\rightarrow
\mathrm{Context}
\rightarrow
\mathrm{Action}.
\]

For clarity, the masks in this appendix are binary visibility masks.
For any mask $M^{*}_{q,k}\in\{0,1\}$,
$M^{*}_{q,k}=1$ indicates that query position $q$ is allowed to attend
to key position $k$. We use $\mathbb{I}[\cdot]$ to denote the indicator
function.

\paragraph{Event-level mask.}
Cross-event attention follows temporal causality. A token may attend to
all tokens belonging to earlier events, but not to future events:
\begin{equation}
M^{E}_{q,k}
=
\mathbb{I}[e_k < e_q].
\label{eq:multi_event_mask}
\end{equation}

\paragraph{Exposure-level mask.}
Different exposures within the same event correspond to co-occurring
candidate items. Since these candidates do not have a temporal ordering,
tokens from one exposure cannot attend to tokens from another exposure
within the same event. Visibility within the current event is therefore
restricted to the same exposure:
\begin{equation}
M^{O}_{q,k}
=
\mathbb{I}[e_k=e_q]
\wedge
\mathbb{I}[o_k=o_q].
\label{eq:multi_exposure_level_mask}
\end{equation}

\paragraph{Lifecycle-level mask.}
Within the same exposure, attention follows the lifecycle order. A token
may attend only to lifecycle stages that are available no later than its
current stage:
\begin{equation}
M^{L}_{q,k}
=
\mathbb{I}[c_k\leq c_q].
\label{eq:multi_lifecycle_mask}
\end{equation}

\paragraph{Final attention mask.}
The final visibility relation combines the three constraints:
\begin{equation}
M^{\mathrm{Attn}}_{q,k}
=
M^{E}_{q,k}
\vee
\left(
M^{O}_{q,k}
\wedge
M^{L}_{q,k}
\right).
\label{eq:multi_exposure_attention_mask}
\end{equation}
Thus, a token can attend to all tokens from earlier events. Within the
current event, it can attend only to tokens from the same exposure whose
lifecycle stages do not succeed its current stage. Tokens belonging to
different exposures of the same event remain mutually invisible.

To match the additive attention mask used in the main model, the binary
visibility mask is converted as
\begin{equation}
\widetilde{M}_{q,k}
=
\begin{cases}
0, & M^{\mathrm{Attn}}_{q,k}=1,\\
-\infty, & M^{\mathrm{Attn}}_{q,k}=0.
\end{cases}
\label{eq:multi_exposure_additive_mask}
\end{equation}

When every event contains exactly one exposure, i.e.,
$o_q=o_k$ for tokens within the same event, the exposure-level constraint
becomes trivial. Under the event-major lifecycle serialization used by
LIFT, Eq.~\ref{eq:multi_exposure_attention_mask} then reduces to the
standard lower-triangular causal mask:
\begin{equation}
M^{\mathrm{Attn}}_{q,k}
=
\mathbb{I}[k\leq q].
\label{eq:multi_standard_causal_mask}
\end{equation}

\end{document}